\documentstyle[12pt,aaspp4,epsf]{article}

\def\be{\begin{equation}}
\def\ee{\end{equation}}
\def\bea{\begin{eqnarray}}
\def\eea{\end{eqnarray}}

\def\citeauthoryear#1#2#3{}

\lefthead{Dubinski}
\righthead{Origin of BCG's}

\begin{document}

\title{The Origin of the Brightest Cluster Galaxies}

\author{John Dubinski}
\affil{Canadian Institute for Theoretical Astrophysics\break
    McLennan Labs\break
    University of Toronto\break 60 St. George St.,\break
    Toronto, Ontario M5S 3H8, Canada\break
    dubinski@cita.utoronto.ca}

\begin{abstract}
Most clusters and groups of galaxies contain a giant elliptical
galaxy in their centres which far outshines and outweighs normal
ellipticals.
The origin of these brightest cluster galaxies is intimately related
to the collapse and formation of the cluster.  
Using an N-body simulation of a cluster of galaxies
in a hierarchical cosmological model, we show that galaxy
merging naturally produces a massive, central galaxy with 
surface brightness and velocity dispersion profiles similar to observed BCG's.
To enhance the resolution of the simulation, 100 dark halos at $z=2$
are replaced with self-consistent disk+bulge+halo galaxy models following
a Tully-Fisher relation using 100000 particles for the 20 largest galaxies
and 10000 particles for the remaining ones.  This technique allows us to
analyze the stellar and dark matter components independently.
The central galaxy forms through the merger of several massive galaxies 
along a filament early in the cluster's history.  Galactic cannibalism 
of smaller galaxies through dynamical friction over a Hubble time 
only accounts for a small fraction of the accreted mass. 
The galaxy is a flattened, triaxial object whose long axis aligns 
with the primordial filament and the long axis of the 
cluster galaxy distribution agreeing with observed trends for galaxy-cluster
alignment.  
\end{abstract}


\section{Introduction}

The brightest cluster galaxies (BCG's) are the most 
luminous and massive galaxies in the universe.
A typical BCG is located  near the centre of its parent cluster
and well-aligned with the cluster galaxy distribution suggesting that it
lies at the bottom of the cluster's gravitational potential well.
The general impression that stars have settled to the bottom of a pit
suggests that the origin of BCG's is closely connected to the formation
of the cluster.
BCG's are elliptical galaxies that are
much brighter and much more massive than the average with
luminosities 
$\sim$10 L$_*$ ($L_* = 1.0\times 10^{10} h^2$
L$_\odot$),
(e.g., Sandage \& Hardy 1973; Schombert 1986; Brown 1997)
\cite{Sandage73}\cite{Schombert86}\cite{Brown97}
central velocity dispersions 
in the range $\sigma \sim 300-400$ km/s 
(e.g., Dressler 1979; Carter et al. 1985; Fisher,
Illingworth \& Franx 1995)\cite{Dressler79}\cite{Carter85}\cite{Fisher95}
and very little rotational support. 
Like other ellipticals, their light profile is well-described by a
deVaucouleurs surface brightness law, $\mu(r) \propto r^{1/4}$
over a large range in radii (deVaucouleurs 1948)\cite{deVaucouleurs48}.
The BCG's are variously
classified as giant ellipticals (gE),
as D galaxies which have somewhat shallower light profiles 
than E's and the final classification cD for D galaxies with an 
extended envelope of excess light over and above a deVaucouleurs 
law fit to the inner regions (Kormendy 1989)\cite{Kormendy89a}.
cD galaxies are also only found in the centres of clusters and groups 
so their extended envelope is probably associated with the 
formation of the cluster.

The following theories have been proposed to explain the origin of BCG's,
i) star formation from cooling flows
expected in the high density, rapidly cooling centres of cluster X-ray 
halos (Fabian 1994)\cite{Fabian94},
ii) galactic cannibalism or
the accretion of the existing galaxy population through dynamical 
friction and tidal stripping (Ostriker \& Tremaine 1975; Richstone 1976;
Ostriker \& Hausman 1977)\cite{Ostriker75}\cite{Richstone76}\cite{Ostriker77} and 
iii) galaxy merging in the early history of the
formation of the cluster as expected in hierarchical cosmological
models (Merritt 1985; Tremaine 1990)\cite{Merritt85}\cite{Tremaine90}.
The cooling flow theory implies the creation of lots of new stars but
generally there is weak evidence for this
population (McNamara \& O'Connell 1989)\cite{McNamara89}.
The galactic cannibalism picture fails when worked out in detail since the
dynamical friction timescales are generally too  long 
and so the expected amount of accreted luminosity  falls short 
by an order of magnitude for making up a BCG's 
luminosity (Merritt 1985; Lauer 1985; Tremaine 1990)\cite{Merritt85}\cite{Tremaine90}\cite{Lauer85}.
The failure of this model implies that BCG's must have an earlier origin
and that galaxy merging within the cluster 
during collapse in a cosmological hierarchy is a possible alternative.
The strong tendency for BCG's to align with their cluster 
population (Sastry 1968; Carter et al. 1980; Binggeli 1982; West 1994)\cite{Sastry68}\cite{Carter80}\cite{Binggeli82}\cite{West94} also implies 
an origin coinciding with cluster collapse.

Most of the work on the formation of giant ellipticals has been
based on studies of merging groups of several disk
galaxies (Barnes 1989; Weil \& Hernquist 1996) or small virialized clusters
of spherical galaxies (Funato, Makino \& Ebisuzaki 1993; Bode et al. 1994;
Garijo, Athanassoula \& Garcia-Gomez 1997).  These simulations reveal the
high efficiency of dynamical friction in driving galaxy merging and
the general tendency to produce remnants resembling elliptical galaxies. 
However, they are phenomenological studies that are still considerably
detached from the context of hierarchical collapse
in which elliptical galaxies and BCG's probably form.
In this paper, we explore galaxy merging in a detailed cosmological
simulation of cluster collapse including a realistic distribution
of disk galaxies embedded in dark halos and show that it produces a consistent
and quantitative picture for the origin of BCG's.

\section{Merging and the Formation of Elliptical Galaxies}

The formation histories of BCG's and ordinary elliptical galaxies are
closely linked.
Elliptical galaxies most likely form through the dissipationless merger of
smaller spiral or elliptical galaxies (Toomre \& Toomre 1972; Toomre 1977),
\cite{Toomre72}\cite{Toomre77} while the
spirals themselves form dissipatively as gas cools radiatively and sinks to
the centre of a dark halo (White \& Rees 1978)\cite{White78}.
When galaxies collide, dynamical friction combined with
strong time-dependent mutual tidal forces 
redistribute the ordered orbital kinetic energy into random energy allowing the galaxies to merge into an amorphous, triaxial system resembling an
elliptical galaxy 
(Barnes 1988; Hernquist 1992,1993; Barnes \& Hernquist 1992ab)
\cite{Barnes92b}\cite{Barnes88}\cite{Barnes92a}\cite{Hernquist92}\cite{Hernquist93}.
In the various N-body studies of mergers of galaxy pairs,
analysis of the merger remnants show that they have 
light profiles and kinematics similar to observed ellipticals, 
although the match is not perfect.
The simulation of the merger of groups of galaxies
(Barnes 1989; Weil \& Hernquist 1996)\cite{Barnes89}\cite{Weil96}
which is more closely related to the formation of BCG's
give generally similar results to the merger of galaxy pairs.
Simulated merger remnants generally have deVaucouleurs profiles, although
the cores can be less dense than real ellipticals when the progenitors are
pure stellar disks without a bulge (Hernquist 1992)\cite{Hernquist92}.
This problem probably arises from Liouville's theorem and the
conservation of fine-grained phase
space density (Carlberg 1986)\cite{Carlberg86}. 
The central phase space density of the
remnant can be no greater than that of the progenitors 
so pure disks with low central phase-space density cannot make ellipticals
with a high central value.
Disk galaxies with bulges can lead to denser cores (Hernquist 1993)\cite{Hernquist93},
but this only skirts the issue by including
an elliptical component with a dense core in the progenitors.
The solution to this problem is
probably gaseous dissipation through radiative cooling 
which can lead to higher, central densities in merging galaxies with 
gas (Kormendy 1989b)\cite{Kormendy89b}, 
although simulations with gas and star formation seem
to produce cores which are too dense (Mihos \& Hernquist 1994)\cite{Mihos94}.
Despite the uncertainties in core properties, 
galaxy merging produces remnants with global structure and kinematics 
similar to real ellipticals and remains the most likely way
that they form.

Galaxy clusters hold the key to understanding the formation of
elliptical galaxies.  While E galaxies only make up about 10\% of
all galaxies,  they are much more abundant in regions of high galaxy
density, especially in the centres of clusters of galaxies where they make
up most of the galaxy population 
(Dressler 1980; Dressler 1984)\cite{Dressler80}\cite{Dressler84b}.
The high frequency of E galaxies in rich clusters has been viewed as a
paradox and evidence against the merger hypothesis, 
since the large relative velocities of galaxies in {\em virialized}
clusters should not permit galaxy merging (Ostriker 1980)\cite{Ostriker80}.
The high number density of galaxies may permit the merger of some galaxies in
the low velocity tail of a virialized distribution (Mamon 1992) 
that may account for the observed elliptical concentration,
but this picture does not include the effects of hierarchical collapse.
In a cosmological hierarchy, small groups of galaxies will form
prior to the collapse and virialization of the cluster and the velocity
dispersion in these groups may be low enough to permit dissipationless
merging and the formation of elliptical galaxies.
Small elliptical galaxies may also be created in gravitational
instabilities in the tidal tails
of interacting galaxies (Barnes 1992)\cite{Barnes92} or from
galaxy harassment, the cumulative tidal perturbations 
from other galaxies or the cluster potential 
(Moore et al. 1996a)\cite{Moore96a}.
Disk galaxies need a quieter environment to form through
dissipative collapse and the centre of a cluster where strong tidal forces
from closely interacting protogalaxies is probably the least likely
place to form a disk.  One might therefore expect elliptical galaxies and in
particular a BCG to form in the cluster centre where the lumpy
mass flow is converging.  

While phenomenological studies of galaxy mergers are useful 
for establishing generic properties of merger remnants,
the initial trajectories are only roughly based on cosmological expectations
and the galaxy dark halo models are usually truncated at a smaller mass 
and extent than seen in halos in cosmological 
simulations 
(Dubinski \& Carlberg 1991; Navarro, Frenk \& White 1996).
\cite{Dubinski91}\cite{Navarro96}
An examination of galaxy merging in the cosmological setting
of a cluster is therefore necessary to go beyond these studies,
though,
the large dynamic range in mass 
between galaxies and clusters (a factor of 1000 or more)
make cluster simulation with resolved galaxies 
difficult to study using either N-body or combined N-body/gasdynamics methods.
Galactic dark halos seem to merge too efficiently and form 
a smooth cluster dark halo with very little internal substructure 
corresponding to galaxies at late times (Moore et al. 1996b)\cite{Moore96b}.
This problem originates
from inadequate dynamic range from large
gravitational softening lengths (50 kpc) and too few particles.
Larger cluster simulations with $\sim 10^6$ particles show that
substructure corresponding to a galaxy scale can survive during the
formation of a cluster (Carlberg 1994)\cite{Carlberg94}.
In N-body gasdynamical simulations,
galaxies form in the centre of dark halos through gaseous dissipation.
These tightly bound objects may survive infall into the cluster,
but in practice,
the number of galaxies formed in this way in
simulations is highly sensitive to the cooling rate assigned to the 
gas (Frenk et al. 1996)\cite{Frenk96}.
Gasdynamical simulations are also much more computationally intensive
making this method very difficult at present.
%

\section{Simulating Galaxies in a Cluster Collapse}
 
In this paper, we introduce a new approach to cluster simulation 
with a large enough dynamic range to resolve galaxies
within a cluster and examine galaxy merging in a cosmological context.
We simply assume that disk galaxies form instantly in the centre
of galactic-mass dark halos early on in the evolution of the dark matter
cluster in a cosmological N-body simulation.
At an early time, we replace galactic dark halos with
equilibrium, N-body galaxy models scaled to the appropriate mass and
dimensions and resolved with 10-100 times  as many particles. 
We assume that the first galaxies are disks embedded in 
dark halos with flat rotation
curves similar to the Milky Way and other nearby galaxies 
(Kuijken \& Dubinski 1995)\cite{Kuijken95}. 
To guarantee that the chosen galaxies end up in the final cluster we use
the following procedure: 

\begin{enumerate}
\item A dark matter simulation of the cluster is run until the
current epoch, $z=0$.
\item The particles in the final virialized cluster are labelled and
identified at the earlier epoch, $z=2$.
\item From this subset of particles, all galactic dark halos are identified
using the friends of friends linking method 
(Davis et al. 1985)\cite{Davis85} and
replaced by randomly-oriented disk galaxy models 
of the same mass scale placed on the
same trajectories as the halos.
\item The simulation of the cluster is then rerun with the higher
resolution galaxy population until the present epoch.
\end{enumerate}

A similar method has been used in
the study of galaxy harassment in clusters (Moore et al. 1996a)\cite{Moore96a}.
The idea is to
enhance the resolution of the simulation selectively 
by using galaxy models as in phenomenological
studies while retaining the cosmological character of the mass distribution
and large-scale kinematics.


The experimental cluster is chosen from a cold dark
matter (CDM) simulation
of periodic cubic volume with $L=32$ Mpc on a side, assuming $H_o = 50$
km/s/Mpc and normalized to $\sigma_8 = 0.7$.  
With this normalization, the CDM model is a good description of
the clustering of galaxies for the scale we are examining, although it is
known to lack sufficient power on larger scales.
The initial conditions were
generated by applying the Zel'dovich approximation to
a random realization of a CDM density field generated with a $256^3$
Fourier transform.
The simulation is first run at $64^3$ resolution to identify the site of
cluster formation.  A spherical volume of comoving radius $R = 11$ Mpc,
associated with the virialized cluster,
is identified in the initial conditions and then resampled at the full
$256^3$ resolution.  The tidal boundary of the collapsing cluster is
adequately handled using two concentric shells in the radial ranges of
$11<R<16$ Mpc and $16<R/L<27$ Mpc, sampled at $128^3$ and $64^3$
resolution respectively.  The simulation therefore contains a total of 4.3
million particles of which about 1 million end up in the virialized
cluster halo.  All of the simulations are run with a
parallel N-body treecode adapted for both periodic and vacuum 
boundaries (Dubinski 1996)\cite{Dubinski96}.

The cluster has a virial radius of 1.2 Mpc, a mass of $1.0
\times 10^{14}$ M$_\odot$ within this radius and a spherically averaged,
central line-of-sight velocity dispersion of 550
km/s. This cluster would be classified as a poor cluster or a large group
by observers.
At $z=2$, the 100 most massive dark halos associated with the cluster
with masses in the range of $10^{10}$ 
to $5 \times 10^{12}$ M$_\odot$ are identified and replaced
with N-body galaxy models.
The model used for each galaxy (with different scaling) 
is composed of an exponential disk, a truncated King model
bulge and King model dark halo with the potential and orbital distribution
derived from a self-consistent distribution function 
(Model B of Kuijken \& Dubinski 1995)\cite{Kuijken95}.
By design, the model
has a flat rotation curve out to 10 exponential scale-lengths and declines
beyond that distance.

The 20 most massive halos are replaced with ``high'' resolution galaxy
models including 50000 disk particles, 10000 bulge particles and 40000 dark
halo particles with a softening length of 0.32 kpc for the stars and 0.64
kpc for the dark halo particles.
The remaining 80 halos are sampled at ``low'' resolution
with one tenth as many particles as above 
in the same ratios and a softening length twice as large.
The remaining cluster dark matter is retained with a softening length of
3.2 kpc.

\begin{figure}
\centerline{\epsfbox{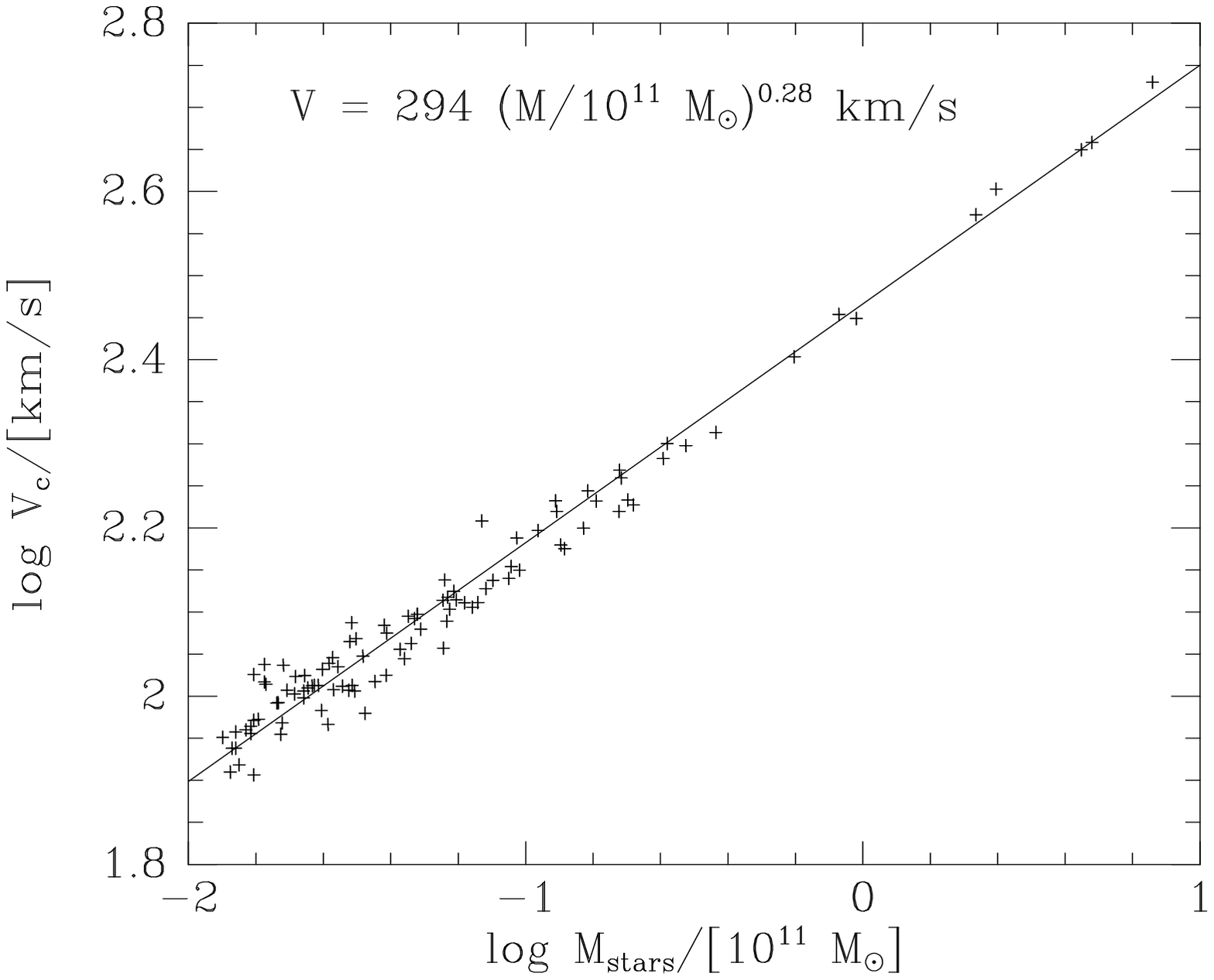}}
\caption{Relation between mass and circular velocity for the initial galaxy
population. The relation is similar to the Tully-Fisher relation.}
\label{fig-mv}
\end{figure}

\begin{figure}
\centerline{\epsfbox{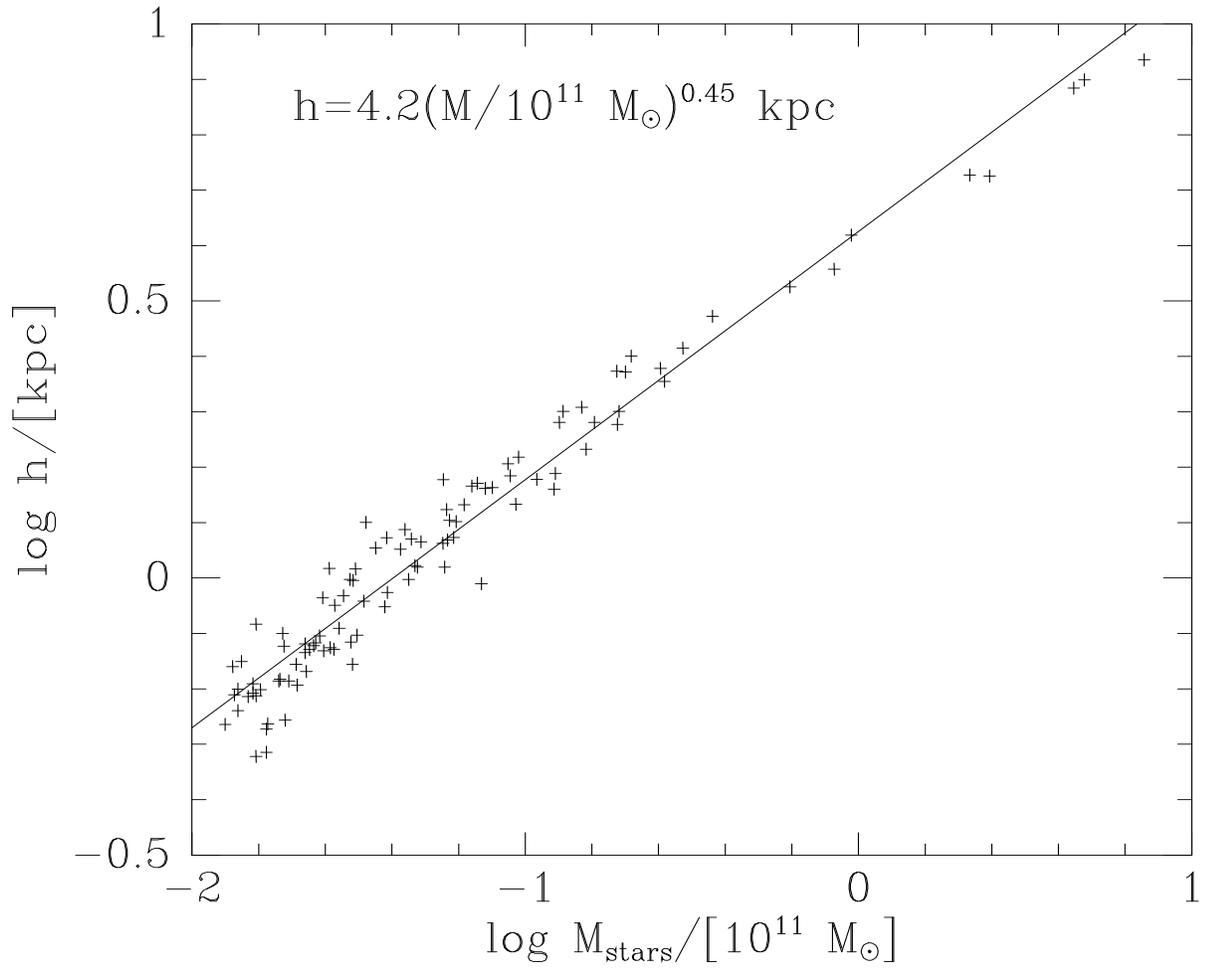}}
\caption{Relation between mass and exponential scale-length for the initial
galaxy population.  The relation follows the prediction of 
Freeman's (1970) law for exponential disks.}
\label{fig-mr}
\end{figure}

The models are scaled according to the value of the circular velocity and
mass of the halos in the dark matter simulation at
about 1/2 the virial radius.
The scale-lengths, $h$, of the disks are determined by the measured
mass and velocity ($h \propto GM/v^2$) and fall in the range of observed
disks.
The 100 galaxies roughly follow a Tully-Fisher relation 
(Tully \& Fisher 1977)\cite{Tully77} 
in their circular
velocity vs. mass profiles with $v_c \sim M^{0.28}$ (Figure \ref{fig-mv}).
The disk scale-lengths
also vary according to observed laws with $h \sim M^{0.45}$ again cf. 
with $h \sim L^{0.5}$ (Freeman 1970)\cite{Freeman70} (Figure \ref{fig-mr}).
The mass function also has a Schechter form with $\alpha \approx -1.5$ 
and $M_* = 5\times 10^{11}$ M$_\odot$ (stellar mass) similar 
to the observed local galaxy luminosity functions but perhaps somewhat 
steeper (Loveday et al. 1992)\cite{Loveday92}.
The simulation is run for 10.5 Gyr with 
a single leapfrog timestep $\Delta t = 2.3$ Myr for a total of 4700 steps.  
This timestep allows the resolution of structures down to about 0.5 kpc.

One problematic feature of the distribution is that the 3 most massive
galaxies have unusually large circular velocities for normal disk galaxies,
with $v_c > 400$ km/s.  These galaxies are probably ellipticals rather than
disks at the time they are selected.  Their exact morphology probably makes
little difference to the final outcome since they quickly merge at the
outset.  The simulation should eventually be rerun with elliptical 
galaxies to check for possible discrepancies.
However, the remaining galaxies follow
a realistic mass distribution having
properties similar to the observed high surface brightness disks.

\section{Results}
\subsection{The Formation of a Giant Elliptical Galaxy}

\begin{figure}
\epsfxsize 4.5in
\centerline{\epsfbox{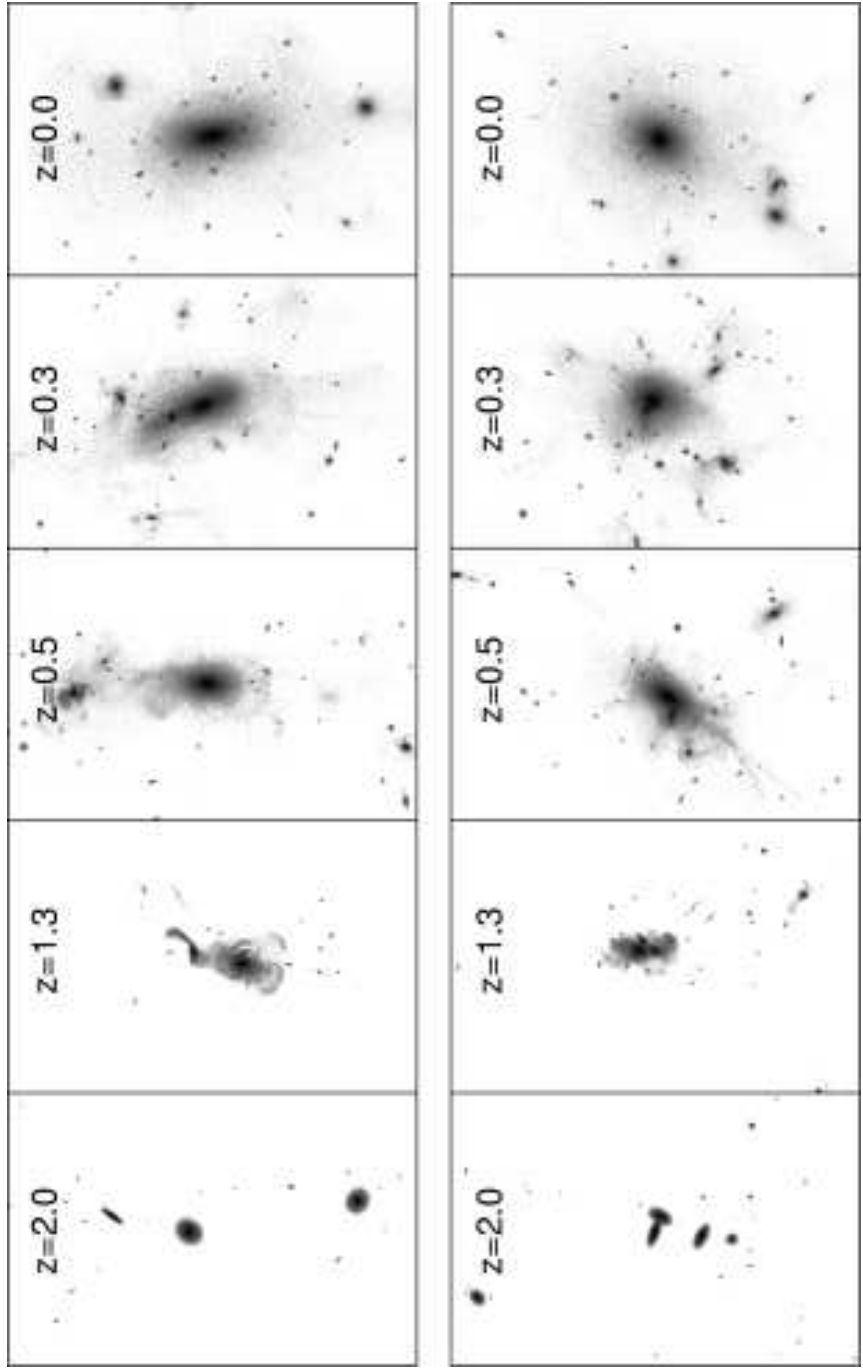}}
\caption{Snapshots of the evolution of the cluster and bright central
galaxy.  Each strip is 1 Mpc wide.  The top strip shows the view
perpendicular to the chain of 3 galaxies which fall together to make the
BCG.  The bottom strip show the view looking approximately down the filament. 
See www.cita.utoronto.ca/$\sim$dubinski/bigcluster.html
for mpeg animations.}
\label{fig-1}
\end{figure}

\begin{figure}
\centerline{\epsfbox{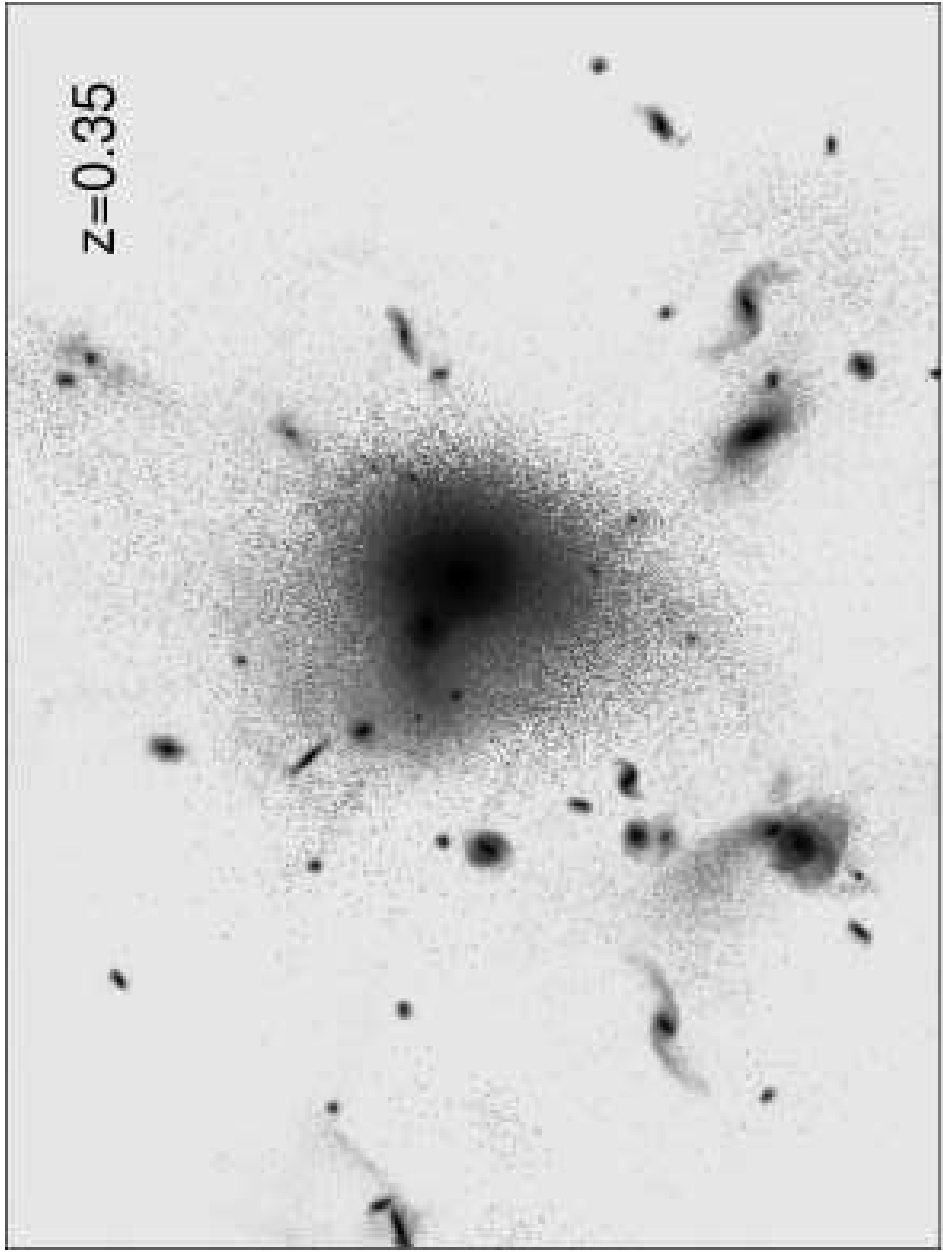}}
\caption{Close-up of the central region of the cluster at $z=0.35$, 
during a very active phase of the collapse.  The giant elliptical is
undergoing a major merger while various disk galaxies are throwing 
off tidal tails resulting from strong tidal interactions with the cluster
centre.}
\label{fig-active}
\end{figure}

Within 3 Gyr of the start of the simulation (by $z=0.8$), 
the four most massive galaxies merge to form a central
object resembling an elliptical galaxy and as we describe below its measured
surface brightness and velocity dispersion profile are very similar to real
giant ellipticals.
Three of the galaxies fall down a line which can be identified
with the primordial filament apparent in the early formation of the dark
matter cluster. The fourth galaxy comes from a different direction, but the
infall of material generally follows the line of the filament 
(Fig.~\ref{fig-1}).
Over the next 5 Gyr (ending around $z=0.4$), 
9 more galaxies are accreted (some as merged products themselves) and of
the total of 13 merging to form the elliptical, 7 are galaxies with circular
velocities greater than 200 km/s while the remaining 6 are
smaller galaxies with circular velocities $\sim 100$ km/s.
The epoch around $z=0.4$ is marked by a period of intense activity
in which many of the galaxies are merging and experiencing strong tidal
perturbations resulting in tidal tails from close passes with the 
cluster center (Fig \ref{fig-active}) and is an illustration of the
galaxy harassment process (Moore et al. 1996).
From $z=0.4$ to the present, there are no more large mergers with the BCG.
At the end of the simulation, only 59 galaxies can be identified orbiting
in the cluster.  Of the 41 ``missing'' galaxies, 
13 have merged to form the central massive elliptical galaxy.
The remaining 28 have been incorporated into other galaxies through
mergers.
The group simulated here is too small to detect the density-morphology
effect (Dressler 1984)\cite{Dressler84b},
the cluster centre), 
however, the few ellipticals created are close
to the cluster centre ($R < 500$ kpc) and on eccentric orbits that 
are measureably decaying by dynamical friction.

\subsection{Analysis of the Merger Remnant}

Is the merger remnant in the centre of the cluster a giant elliptical or cD
galaxy?
We measured
the shape of the central galaxy, the surface density and 
velocity dispersion profiles along different lines of sight to answer this.
We also measured the three-dimensional structure and kinematics of the
stars and dark matter for comparison to interpretations of the projected
data.

\subsubsection{Shape}
\begin{figure}
\epsfxsize 2.0in
\centerline{\epsfbox{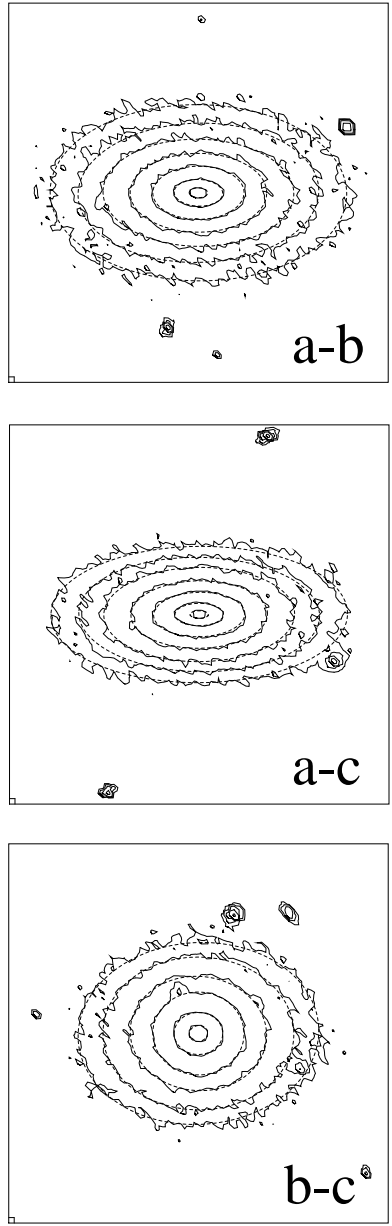}}
\caption{The shape of the central elliptical galaxy viewed down the three
principal axes.  Solid lines are the measured isodensity contours and
dashed lines are the fitted ellipses.}
\label{fig-2}
\end{figure}

The central elliptical is a triaxial object with
principal axis ratios of $b/a=0.66$, and $c/a=0.47$, where $a, b,$ and $c$
are the major, middle, and minor axes (Fig.~\ref{fig-2}).
The surface density contours (isophotes) are nearly perfect ellipses in the
inner regions with very little signal of ``boxiness'' or ``diskiness'' seen
in smaller ellipticals.  The regularity of the isophotes is consistent with
a relaxed galaxy which has suffered no recent mergers, which is indeed the
case.

\subsubsection{Density}

\begin{figure}
\epsfxsize 4.0in
\centerline{\epsfbox{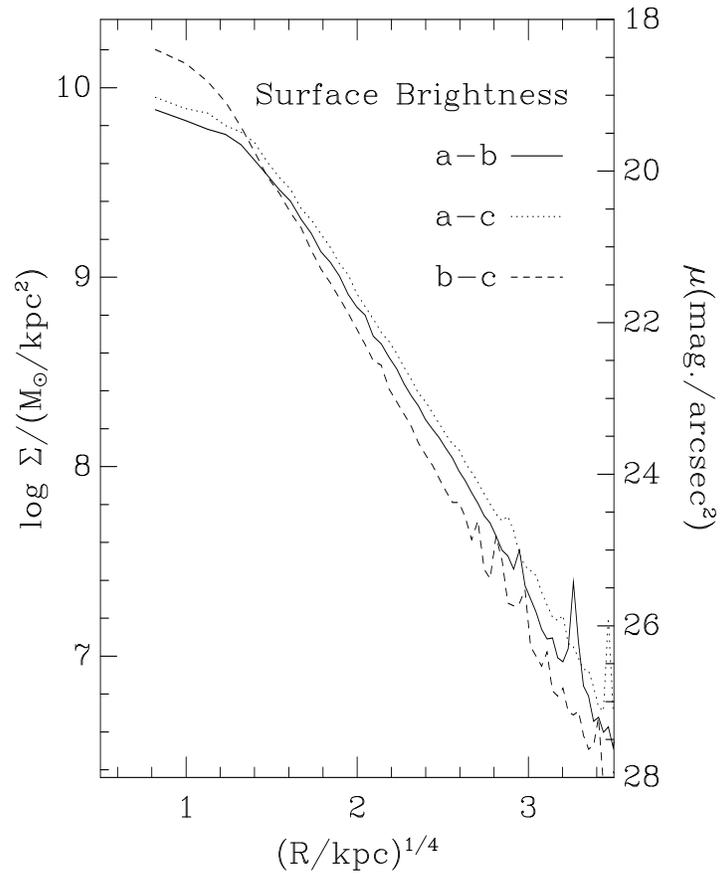}}
\caption{The surface brightness profile of the galaxy along
elliptical isophotes plotted versus $r^{1/4}$ for the principal axis
projections of the galaxy.
The simulated galaxy follows a deVaucouleurs law closely and would be
classified as a giant elliptical.  A cD galaxy envelope did not form in
this simulation.}
\label{fig-3}
\end{figure}

The surface density profiles from three different lines
of sight along each of the principal axes are measured by fitting
elliptical contours to the observed density map.
Figure~\ref{fig-3} 
shows the log surface density $\mu$ plotted versus $r^{1/4}$.  Another 
axis is included showing the equivalent surface brightness assuming
$M_\odot/L_\odot = 10$ for the stars.
The nearly linear dependence on $r^{1/4}$ from 3-100 kpc reveal
that the galaxy follows a deVaucouleurs light profile.  The profile is
similar to measured profiles of giant ellipticals (gE) rather than cD
galaxies which show an excess of light at large radii.
deVaucouleurs profiles are fit
for the effective radius, $r_e$, and the total mass.
The effective radii (calculated as $\sqrt{ab}$) is between 18-22 kpc
depending on the projection.
The fitted total mass is $2.6\times10^{12}$ M$_\odot$
corresponding to a total luminosity of 6 $L_*$ (assuming
$M_\odot/L_\odot = 10$ and $L_* = 1.0\times 10^{10} h^{-2}$ L$_\odot$).  
The regularity of the isophotes and the measured scales and luminosities
of the merger remant are consistent with observations of giant elliptical
galaxies.

\begin{figure}
\epsfxsize 4.0in
\centerline{\epsfbox{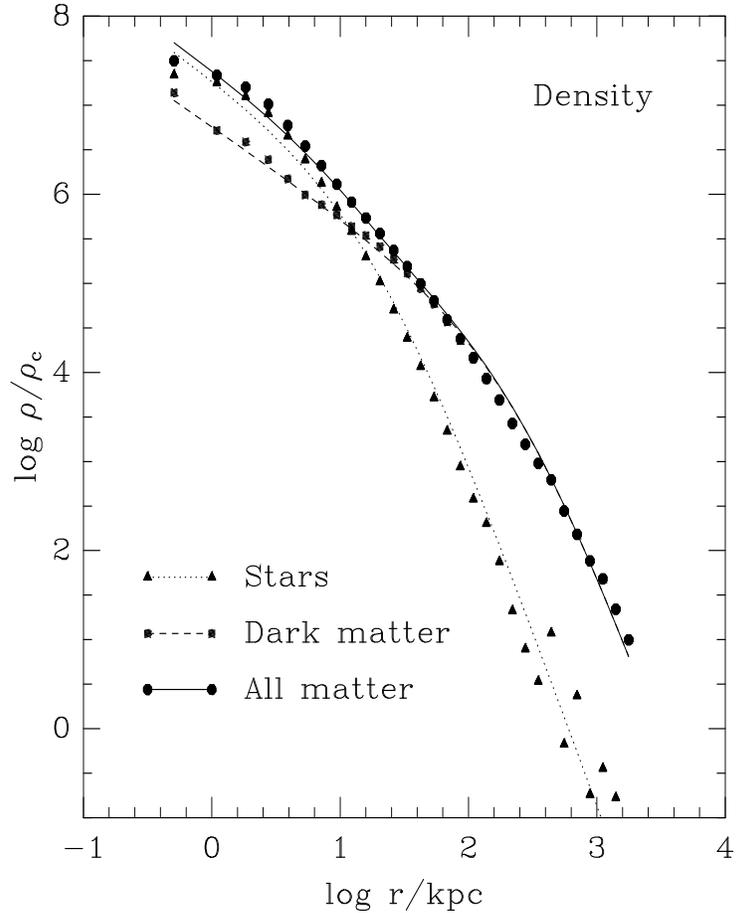}}
\caption{The spherically averaged density profile for the stars, dark
matter and the all of the matter.  The Hernquist profile is fitted
separately to the stars and dark matter profile and the lines are shown.
The solid line is the sum of the two fits.  The stellar density is about 3
times the dark matter density within 10 kpc or 0.5 $r_e$ and so central
kinematics are dominated by the observed stars.}
\label{fig-den}
\end{figure}

The 3-dimensional density profile of the merger remnant reveals the
relative distribution of stars and dark matter in an elliptical galaxy (Fig.
\ref{fig-den}).  The Hernquist density profile (1990)\cite{Hernquist90},
\begin{equation}
\rho(r) = \frac{Ma}{2\pi}\frac{1}{r(a+r)^3}
\end{equation}
provides convenient model
fits to both the stellar and dark matter profiles. The fitted masses and scale
radii are $M_* = 3.3\times 10^{12}$ M$_\odot$ and $a_*=19$ kpc for the stars
and $M_d = 1.6\times 10^{14}$ M$_\odot$ and $a_d = 250$ kpc for the dark
matter as shown in Figure \ref{fig-den}.
Stellar mass dominates within $r<10 kpc$ ($0.5 r_e$),
although the stellar density is only about 3 times the dark matter density
at the center.
The stellar and
dark matter density are equal in the range of 10-20 kpc ($0.5-1.0 r_e$)
while beyond 40 kpc ($2 r_e$) the dark matter density is at least ten times the
stellar density.  The merging process tends to enhance the ratio of dark mass
to luminous mass within in the central regions.  In the initial population
of disk+bulge galaxies the dark to luminous mass ratio at the half light
radius is 0.4, while the giant elliptical has a ratio of about 1.0 at a
nominal effective radius of 20 kpc, a factor of 2.5 enhancement.
The most likely reason for this enhancement is the tendency for disk stars
to be heated more effectively than the dark matter particles.  
The same resonant interactions which create tidal tails during mergers
add energy more effectively to disk stars than to dark matter particles with
the same initial binding energy.  The dark matter density may then be
enhanced slightly with respect to the stars in the merger remnant in
comparison to the initial disks.  

\begin{figure}
\epsfxsize 4.0in
\centerline{\epsfbox{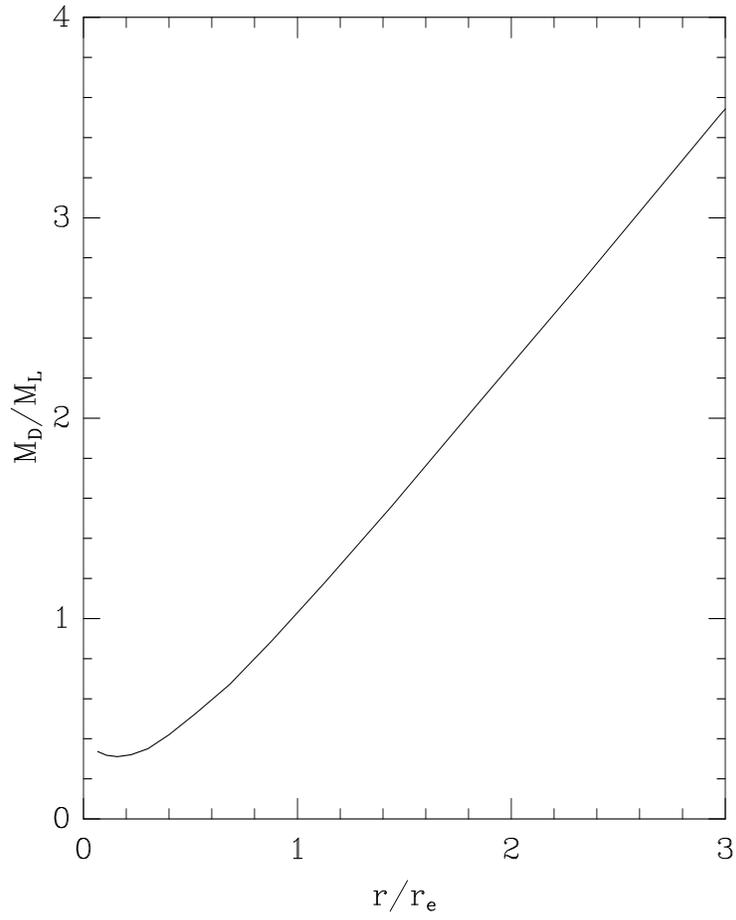}}
\caption{The ratio of dark to luminous mass in the BCG over 5$r_e$.
Dark mass still represents about 30\% of the mass at the center in this
model.  The ratio grows almost linearly with radius.}
\label{fig-mass}
\end{figure}

The overall trend for increasing dark to luminous mass ratio is shown in
Figure \ref{fig-mass}.  Stars dominate the central density
and the dark to luminous mass ratio is only about 0.3.  At $r=0.5 r_e$,
the ratio starts to grow linearly reaching 1.0 at $r=r_e$ continuing
to rise to 3.3 at $r=3 r_e$.  These mass ratios and general behaviour are
in accord with recent models of the dark matter in ellipticals derived from 
combining surface brightness and kinematical information (e.g. Saglia et
al. 1992, 1993; Rix et al. 1997)\cite{Saglia92}\cite{Saglia93}\cite{Rix97}.

\subsubsection{Kinematics}

\begin{figure}
\epsfxsize 4.0in
\centerline{\epsfbox{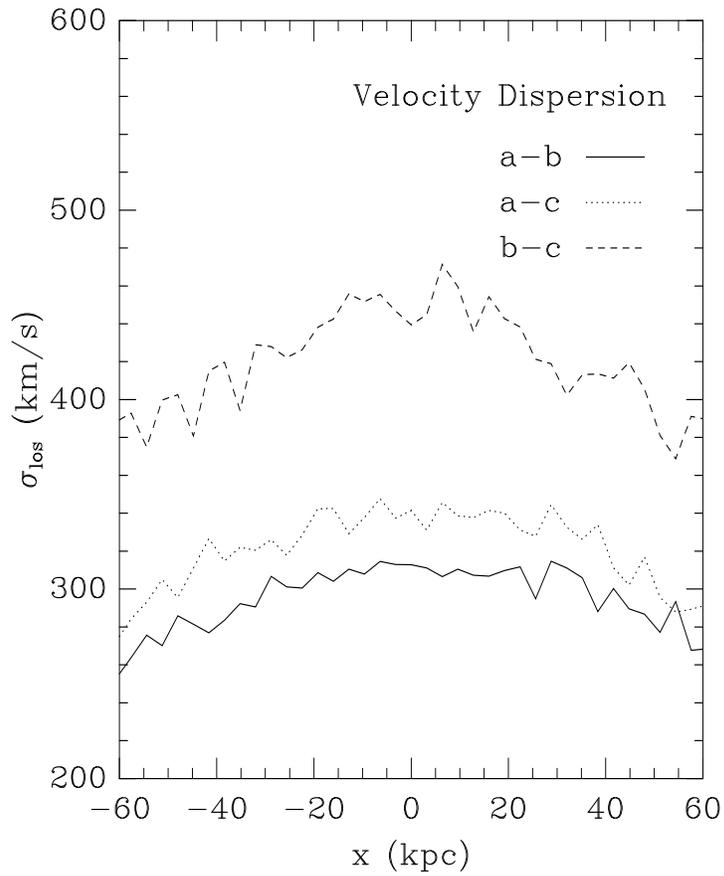}}
\caption{Velocity dispersion profile measured along a slit laid on the
major axis for the three principal axis projections of the galaxy.  
The velocity dispersion 
declines slowly with distance from the centre.  There is no sign of an
upturn at large distances.}
\label{fig-4}
\end{figure}

The velocity dispersion profiles of the galaxies are also measured using a
method faithful to current observational techniques.  A slit 3.2 kpc in
width
is laid along the apparent major axis in three independent directions.
This corresponds observationally to a 1.5 arcsecond slit laid across a
galaxy at a distance of 100 Mpc.
Particles are binned in squares 3.2 kpc on a side and the mean line-of-sight
velocity and velocity dispersion is measured in each bin.
Like real ellipticals,
the galaxy rotates slowly about its minor axis 
with $v_{rot} \sim 50$ km/s (Franx, Illingworth \& Heckman 1989)\cite{Franx89}.
Figure~\ref{fig-4} 
shows the velocity dispersion profile along the apparent major
axis for the three lines of sight down each of the principal axes.
The central value peaks between 300 and 450 km/s depending on the line of
sight.  The large value of 450
km/s occurs when looking exactly down the long axis of the galaxy 
showing the anisotropy of the velocity ellipsoid in this flattened triaxial
stellar system.  These central values again are in accord with real giant
ellipticals although the value of 450 km/s might be considered
too large (Fisher et al. 1995)\cite{Fisher95}.
The velocity dispersion only declines gradually out to 60 kpc (about 3
effective radii) again in similar fashion too many elliptical galaxies.
There is no sign of an upturn in the velocity dispersion at large radii
as seen in the exceptional case of the cD galaxy in A2029
(Dressler 1979)\cite{Dressler79}.

\begin{figure}
\epsfxsize 4.0in
\centerline{\epsfbox{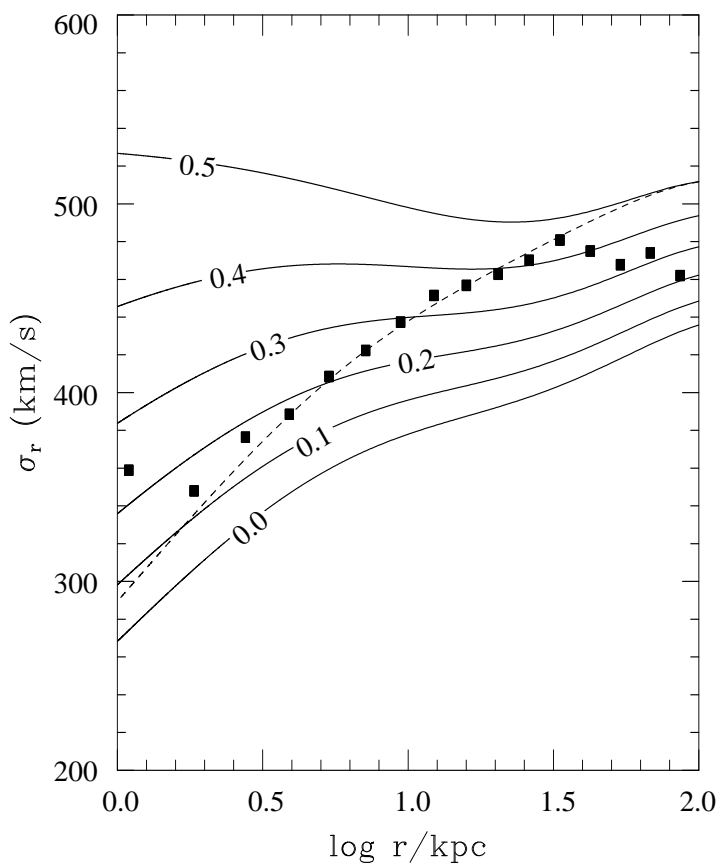}}
\caption{The spherically averaged radial velocity dispersion profile compared
to anisotropic spherical model predictions from the fitted density profile.
Each line is labelled with the anisotropy parameter used in the model.
The anisotropy of the model rises from about 0.2 in the center to 0.5 at
100 kpc ($5 R_e$).  The dashed line represents a best fit model with $\beta(r)$
growing monotonically from 0.0 to 0.5 from the center to 100 kpc.}
\label{fig-vsig}
\end{figure}

In three dimensions, the measured velocity dispersion is nearly isotropic to the
centre but becomes radial anisotropic with a radial anisotropy 
parameter (Binney \& Tremaine 1987)\cite{BT87},
$\beta=0.5$ at 3 $r_e$.
The density profiles of Figure \ref{fig-den} for 
the dark matter and stars were fit with Hernquist
(1990) models and used to solve the spherical Jeans
equations for the velocity dispersion profile of the stars
using constant values for the anisotropy parameter, $\beta$.
Figure \ref{fig-vsig} shows that the velocity profile is consistent with
the mass model for values of $\beta < 0.3$ within $r < 20$ kpc ($r < r_e$).
The best fit spherical Jeans model in Figure \ref{fig-vsig} is one 
where the anisotropy grows monotonically from the center with $\beta=0.0$
to $\beta=0.5$ at 3 $r_e$.

\subsubsection{Comparison with other merger remnants}


The properties of the merger remnant in this experiment are rather
different from those found in other simulations of merging groups.
Weil and Hernquist's (1996)\cite{Weil96} simulations
produced remnants which are nearly
oblate with a small flattening.  They also have a modest amount of
rotational support about their minor axes.  
These results contrast with the prolate
nearly non-rotating object in the cluster simulation.
The differences can be attributed to initial conditions.   The group
simulations start with galaxies on random trajectories selected from an 
isotropic distribution while the cosmological mergers are strongly anisotropic
because of the initial filamentary structure.  
The initial isotropy leads to a rounder, oblate remnant while collapse
down a filament appears to lead to a prolate, elongated object.

\subsection{Fossil Alignments}

BCGs are often well-aligned with the
distribution of the cluster galaxies (Sastry 1968; Carter el al. 1980;
Porter, Schneider \& Hoessel 1991)\cite{Sastry68}\cite{Carter80}\cite{Porter91}
as well as more extended large-scale clustering
features (Binggeli 1982)\cite{Binggeli82} 
and it has been suggested that this is due to
filamentary collapse expected in hierarchical structure formed from
Gaussian random noise 
(e.g. Rhee \& Roos 1990; West 1994; Bond, Kofman, \& Pogosyan 1996)\cite{Rhee90}\cite{West94}\cite{Bond96}.
The alignment of the central galaxy is seen as the consequence of an 
anisotropic collapse remembered from the initial random density field.
Filamentary collapse leads to nearly head-on collisions of
galaxies which create prolate merger remnants aligned with the
initial collision trajectory (Villumsen 1982).  

The simulated BCG examined here
shows the alignment effect as seen in previous work.
The shape and orientation of the BCG is indeed nearly 
congruent with the galaxy distribution in the cluster.
The angle between the long axis of the central galaxy and the 
cluster galaxy population as measured by its moments
is only 15$^\circ$.
Furthermore, the orientation of the BCG is closely aligned with the
primordial filament delineated by the 3 large galaxies which make up
most of the mass of the BCG.   
We should emphasize that this alignment effect only
works for the central, giant elliptical where the kinematics and morphology
are dependent on the large-scale convergence of the flow of matter into the
cluster's forming potential well.  Other galaxies falling into the cluster
that avoid merging with the BCG will have random alignments dependent on
their merging history and tidal interactions with the cluster core and
other galaxies.
This simulation strongly supports the 
hypothesis that the shape and orientation of the central galaxy
are fossils of  the filamentary initial conditions of 
the cluster collapse, although further simulations should be done to
confirm this result.

\section{Conclusions}

In summary, when a population of disk galaxies with an observed
distribution of masses falls in a collapsing cluster in a cosmological
setting, a central, giant elliptical galaxy will form in the cluster
center.
The galaxy forms through the merger of
many  smaller galaxies which converge on the cluster center along 
the filamentary structure originating in the initial density field.
The one simulation presented here agrees quantitatively in its structure
and kinematics with many BCG's but there are still many open questions,
in particular:
What is the origin of the envelope in the cD galaxies in rich clusters?  
Perhaps the driving mechanism is tidal stripping 
(Richstone 1976)\cite{Richstone76} and
harassment (Moore et al. 1996a)\cite{Moore96a} 
which only occurs in more massive clusters than
examined here.  Why does BCG central velocity dispersion poorly 
correlate with the cluster velocity dispersion?  
Are multiple nuclei in BCG's ongoing mergers or the result 
of chance projections?
How do the properties and the timing of formation of BCG's 
depend on different cosmological models?
If there is a strong dependence, the observation of their evolution may
constrain cosmological models.
A modest sample of simulations covering different mass scales and cosmological
models can answer these questions quantitatively.

{\bf Acknowledgements}. I thank the Pittsburgh Supercomputing Center for
granting me computing time on the Cray T3D and T3E. 
I also acknowledge useful comments and criticisms from
Michael West, Lars Hernquist,  Scott Tremaine, Ray Carlberg,
Chigurapati Murali, Jame Brown, Howard Yee, Simon White, Gary Mamon and the
referee James Schombert.  Thanks to Joel Welling 
for help in producing animations of this simulation.

\noindent e-mail:dubinski@cita.utoronto.ca

\noindent website:www.cita.utoronto.ca/$\sim$dubinski/

\newpage


\end{document}